\documentclass[aps,prl,reprint,twocolumn,superscriptaddress,groupedaddress,showpacs,
aps,prl,longbibliography]{revtex4-2}

\usepackage[colorlinks=true,  urlcolor=blue,  linkcolor=blue,  citecolor=blue]{hyperref}
\usepackage{graphicx}
\usepackage{amsmath}
\usepackage{amssymb}
\usepackage{bm}
\usepackage{siunitx}
\usepackage{comment}

\newcommand{\ee}{\mathrm{e}}
\newcommand{\ii}{\mathrm{i}}

\newcommand{\wL}{\omega_{\mathrm{L}}}

\newcommand{\kL}{k_{\mathrm{L}}}

\begin{document}
%
\title{Resonant light scattering by a slab of ultracold atoms}
%
%
\author{R.\ Vatr\'e}
\affiliation{Laboratoire Kastler Brossel, Coll\`ege de France, \\ ENS-Universit\'e PSL, Sorbonne Universit\'e, CNRS\\ 11 place Marcelin Berthelot, 75005 Paris, France}
\author{R.\ Lopes}
\affiliation{Laboratoire Kastler Brossel, Coll\`ege de France, \\ ENS-Universit\'e PSL, Sorbonne Universit\'e, CNRS\\ 11 place Marcelin Berthelot, 75005 Paris, France}
\author{J.\ Beugnon}
\affiliation{Laboratoire Kastler Brossel, Coll\`ege de France, \\ ENS-Universit\'e PSL, Sorbonne Universit\'e, CNRS\\ 11 place Marcelin Berthelot, 75005 Paris, France}
\author{F.\ Gerbier}
\affiliation{Laboratoire Kastler Brossel, Coll\`ege de France, \\ ENS-Universit\'e PSL, Sorbonne Universit\'e, CNRS\\ 11 place Marcelin Berthelot, 75005 Paris, France}
\email{fabrice.gerbier@lkb.ens.fr}
\date{\today}
\begin{abstract}
A gas of ultracold atoms probed with laser light is a nearly-ideal experimental realization of a medium of resonant point-like scatterers, a key problem from condensed matter to biology or photonics. Yet, several recent experiments have reported large discrepancies with theory. In this work, we measure the complex transmission through a slab of ultracold two-level atoms with an interferometric technique. We find good agreement with first-principles simulations of mutually-coupled, laser-driven dipoles, and provide an explanation for the discrepancies in earlier measurements.
\end{abstract}
%
\maketitle
%
%

\definecolor{light-gray}{gray}{0.5}

Studying the interaction of various forms of matter (dense or dilute, crystalline or amorphous, etc.) with light has been central to the progress of fundamental physics. 
Linear dielectric media, in particular, are characterized by their dielectric susceptibility $\chi=\chi'+\ii \chi''$. The imaginary part $\chi''$ determines the ``extinction'' of a probe light beam transmitted through the sample, while the real part $\chi'$  determines the refractive index  and the phase shift in transmission. In most gases at room temperature, the dielectric susceptibility is well described by the celebrated Clausius-Mossotti formula, $\chi = \alpha_0 \rho_\mathrm{3D} /(1-\alpha_0\rho_\mathrm{3D}/3)$, with $\rho_\mathrm{3D}$ the atomic density and $\alpha_0$ the single-particle electric polarisability~\cite{FeynmanLop09}. 

Recent studies have pointed out the shortcomings of the Clausius-Mossotti equation for optically dense atomic gases with $\rho_\mathrm{3D} \kL^{-3} \gtrsim 0.1$ (with $\kL$ the probe light wavenumber)~\cite{Ruostekoski97,Chomaz12,Javanainen13,Javanainen17}. Resonant multiple scattering of photons~\cite{Lagendijk96} leads to light-induced dipole-dipole interactions (DDI)~\cite{Chomaz12,Cherroret16,Roof16,Peyrot18}, collective Lamb shifts\,\cite{Friedberg73,Roehlsberger10,Keaveney12,Meir14,Roof16,Peyrot18} and collective radiation responsible for, \textit{e.g.}, super- and sub-radiant decay of excited atomic ensembles~\cite{Dicke54,Kwong14,Araujo2016,Guerin16,Guerin17}. Note that these collective effects are distinct from effects due to the quantum statistics of the constituents particles\,\cite{Morice95}, as recently measured in \cite{Bons16} for a 3D Bose gas near the Bose-Einstein transition temperature.

Collective effects are captured by the so-called coupled dipole (CD) model, where each atom is described as a point-like scatterer fixed at a static (typically random) position and coupled by a dipolar electric interaction to the probe light and to the field created by the other scatterers in the ensemble~\cite{Foldy45,Lax51,Lax52,Lagendijk96}. 
This model has been extensively used to study light scattering from random media~\cite{VanRossum99,Vynck23}. The CD model is in principle well-suited to describe the optical response of a gas of ultracold atoms where atomic motion and the associated Doppler broadening are negligible~\cite{Morice95,Chomaz12,Javanainen13,Bromley16,Javanainen17}. However, several recent experiments~\cite{Jennewein16,Corman17,Jennewein18} measuring the light extinction for moderately dense samples ($\rho_\mathrm{3D}\kL^{-3}\sim 0.1$) disagree with the predictions of the CD model, in some cases even qualitatively~\cite{Corman17}. 
Since the CD model is in fairly good agreement with fluorescence measurements monitoring incoherent scattering~\cite{Pellegrino14,Bromley16,Jenkins16a,Jenkins16b}, the qualitative failure of the same model to account for coherent scattering responsible for extinction has been surprising and remains unresolved. 

In this Letter, we study experimentally the resonant optical response of a quasi-two-dimensional (2D) ultracold atomic gas. Using an interferometric method, we measure both the transmission phase shift and the extinction of a weak probe beam, respectively  related to the real and imaginary parts of the dielectric susceptibility.  We observe a slight distortion of the lineshape but no significant line shift or broadening. The asymmetric lineshape can be explained by a continuous medium approach where atoms are treated as independent scatterers (IS) in a quasi-2D geometry. Our measurements are in good agreement with state-of-the-art numerical simulations of the CD model in the regime of densities explored. 
We point out that in optically dense samples, a direct measurement of the intensity transmission  is easily contaminated by parasitic effects such as off-axis scattering or imaging noise. When accounting for these effects, the apparently mysterious features of light scattering by an ultracold atomic slab disappear. 

\begin{figure}[ht!!!]
\includegraphics[width=\columnwidth]{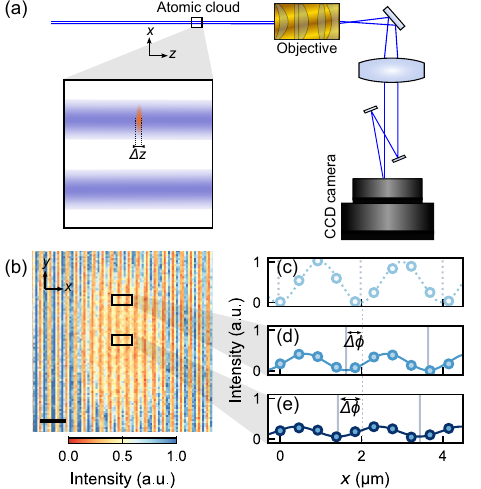}
\caption{Interferometric measurement of the optical response of a slab of ultracold atoms.
(a) Sketch of the experimental setup (not to scale). The inset shows the atomic cloud (r.m.s. width $\Delta z \approx 0.12\,\mu$m) in red  and the probe and reference beams in blue. 
(b) Single-shot interference fringes. (c-e): Illustrative fringe profiles extracted from different regions of the atomic cloud [(d-e), shown by the solid boxes in (b)] or from a  reference region without atoms far from the atomic cloud [c].  
}
\label{Fig:Fig1}
\end{figure}

Experimentally, a thin slab with thickness $\Delta z \sim \kL^{-1}$ offers the highest possible transmission for a given 3D atomic density. In the following, we therefore study  quasi-2D atomic media characterized by a thickness $\Delta z$ and by a normalized surface density
\begin{align}\label{eq:rho2Dnorm}
 \widetilde{\rho}_{\mathrm{2D}} & =  \rho_{\mathrm{2D}} \kL^{-2} ,
\end{align}
with  $\rho_{\mathrm{2D}}$ the atomic surface density. We prepare quantum-degenerate 2D gas of $^{174}$Yb atoms by loading up to $\SI{4.5e4}{}$ atoms in a single site of a large-period ($\approx \SI{4.0}{\micro\meter}$) optical lattice trap in the vertical $z$-direction. 
The atomic density profile along $z$, $f(z)=\ee^{-z^2/(2\Delta z^2)}/\sqrt{2\pi \Delta z^2}$,  is  Gaussian with a root-mean-square (r.m.s.) thickness $\Delta z=\sqrt{\hbar/(2m\omega_z)}\simeq 1.9\,\kL^{-1}$. We illuminate the atomic sample with  circularly-polarized  laser light with a wavevector $\kL \simeq k_0$ and a frequency $\wL $ near-resonant with the transition frequency $\omega_0$ of the ${}^1S_0-{}^1P_1$ transition (wavelength  $2\pi k_0^{-1} \approx \SI{398.9}{\nano\meter}$). The $J=0 \to J=1$ nature of the resonance and the presence of a  magnetic field $\bm B=B_z\bm e_z$\,\cite{SuppMat} ensure that the atoms behave effectively as two-level systems. We keep the probe light intensity $I\approx 0.2\,I_{\mathrm{sat}}$   below the saturation intensity $I_{\mathrm{sat}}\approx 60\,$mW/cm$^{2}$ and illuminate the sample for a duration of $\SI{5}{\micro\second}$ to minimize atomic motion\,(see Supplemental Material for details\,\cite{SuppMat}). 



The coherent optical response of the atomic sample is characterized by a complex transmission coefficient $t(x,y)=\sqrt{T(x,y)}\exp[\ii\Delta\phi(x,y)]$, with $T$ the intensity transmission coefficient  and $\Delta\phi$ the transmission phase shift imprinted on the probe beam by the atomic sample. We measure $T$ and $\Delta\phi$ using the interferometric setup shown in Fig.\,\ref{Fig:Fig1}a. Two mutually coherent beams, a probe beam going through the atomic sample and a reference beam propagating far away from it, are collected by a high-resolution imaging system and recombined with mirrors, such that they overlap with a small angle on a camera to produce interference fringes. Fig.\,\ref{Fig:Fig1}b shows a single-shot image displaying the interference fringes on top of a spatially-resolved image of the density profile of the gas. We define for each image $27$ regions of interest of size $2.3 \times \SI{4.6}{\micro\meter^2}$ with approximately uniform density (we discard realizations where the density varies by more than 10\,\% across the box). Additionally, we also vary the total number of atoms in the cloud  to scan the peak density. Each ``box'' extracted from the collection of recorded images constitutes an experimental realization for a given atomic density. We calibrate independently the atomic density in each region of interest using on-resonance, high-intensity absorption imaging~\cite{SuppMat} where multiple-scattering effects are expected to be highly suppressed~\cite{Hung11a,Yefsah11a}.  The maximum achievable 3D density is $\rho_{\mathrm{3D}}\approx  \SI{3.3e14}{ }$\,at$/\SI{}{\centi\meter^{-3}}=0.09\,\kL^3$, corresponding to a surface density of $\widetilde{\rho}_{\mathrm{2D}}\approx 0.42$.

\begin{figure*}[ht!!!]
\includegraphics[width=\textwidth]{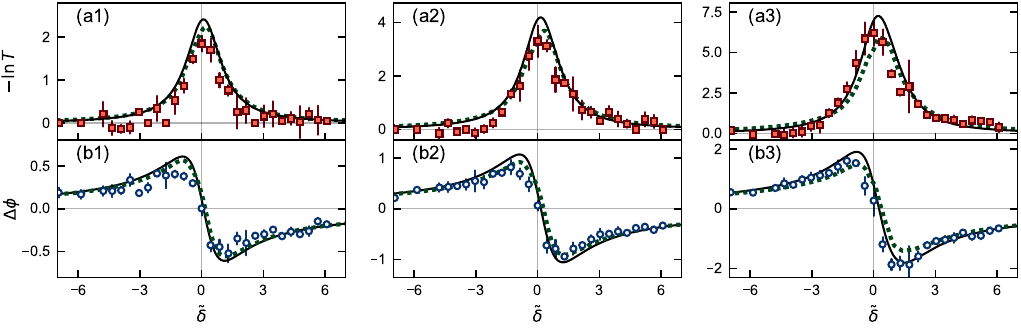}
\caption{Optical response $t=\sqrt{T}\exp(\mathrm{i}\Delta\phi)$ as a function of the normalized detuning $\widetilde{\delta}=2(\omega_\mathrm{L}-\omega_0)/\Gamma_0$. The normalized surface density is (from left to right) $\widetilde{\rho}_{\mathrm{2D}} \approx$ 0.13, 0.23, and 0.42. Solid lines: prediction from a model with independent scatterers taking the Gaussian slab geometry into account. Dotted lines: results from numerical simulations of the coupled dipoles model. The vertical error bars show the standard deviation over three repetitions of the experiment.}
\label{Fig:Fig2}
\end{figure*}

For each experimental realization, we perform fits to the fringe profiles (integrated along $y$) by the sinusoidal function  
\begin{align}
\label{eq:Ifit}
S(x) & =A \big[1+ \gamma \cos(q x +\phi)\big]
\end{align} 
with a wavenumber $q$, a background intensity $A$, a fringe contrast $\gamma $ and a phase $\phi$. We infer the intensity transmission $T$ from the fitted fringe contrast $\gamma \in [0,1]$\,\cite{SuppMat}, and the transmission phase shift $\Delta\phi = \phi_\mathrm{at}-\phi_\mathrm{ref}$  by comparing the phases $\phi_\mathrm{at}$ and $\phi_\mathrm{ref}$ measured respectively in the region containing the atoms and in a reference region away from the atoms (see Fig.\,\ref{Fig:Fig1}c-e). 

\begin{figure}[t!!!]
\includegraphics{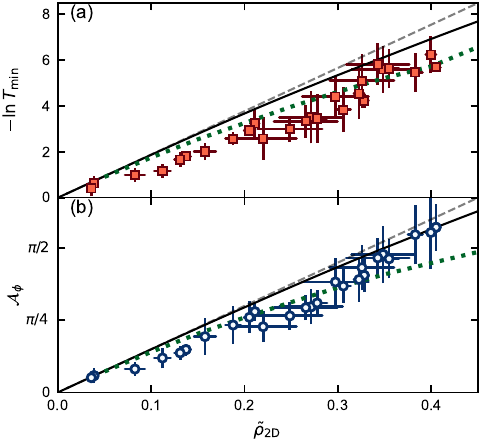}
\caption{(a) Maximum optical depth $-\ln T_{\mathrm{min}}$ and (b) phase shift amplitude $\mathcal{A}_\phi$ as a function of the surface density $\widetilde{\rho}_{\mathrm{2D}}$. Dashed lines: Beer-Lambert prediction for a dilute 3D sample. Solid lines: model of independent scatterers taking the Gaussian slab geometry into account. Dotted lines: numerical simulations of the coupled-dipoles model. Horizontal error bars show the uncertainty inherited from the calibration of the atomic density~\cite{SuppMat}. Vertical error bars are the standard deviation over three repetitions of the experiment.}
\label{Fig:Fig3}
\end{figure}

In Fig.\,\ref{Fig:Fig2}, we show the optical depth $-\ln T$ and the phase shift $\Delta\phi$ as a function of the normalized detuning $\widetilde{\delta}=2(\omega_\mathrm{L}-\omega_0)/\Gamma_0$ (with $\Gamma_0/(2\pi) \approx  \SI{29.1}{MHz}$ the natural linewidth of the transition). For conciseness, we only show data for three representative atomic densities. The maximum  optical depth and phase shift both increase with density as intuitively expected. 
From phenomenological Lorentzian and dispersive fits to the data (not shown, see \cite{SuppMat}), we find neither a statistically significant shift\,\footnote{Ref.~\cite{Corman17} reports a blue shift of the resonance line, but that shift is compatible with zero in the range of densities explored in our experiment.}, nor a broadening of the resonance line~\cite{SuppMat}. Fig.\,\ref{Fig:Fig2} also shows that the resonance line becomes asymmetric  with a skew towards positive detuning as the density increases. This effect, visible by eye on the optical depth curves, is also present (if less noticeable) for the phase shift. We summarize the experimental data in Fig.\,\ref{Fig:Fig3} by plotting the transmission minimum $-\ln T_{\mathrm{min}}$ and the phase shift amplitude  $\mathcal{A}_\phi = [\max_{\widetilde{\delta}} \Delta\phi-\min_{\widetilde{\delta}}\Delta\phi ] /2$ versus the surface density $\widetilde{\rho}_{\mathrm{2D}}$. In both cases, both quantities increase almost linearly with $\widetilde{\rho}_{\mathrm{2D}}$ without sign of saturation.

 The asymmetric lineshape has been proposed as a signature of  DDI (also called ``recurrent scattering'' in the language of multiple scattering theory~\cite{Lagendijk96}), which tend to enhance the transmission when $\widetilde{\delta} >0$ and to suppress it when $\widetilde{\delta} <0$ \,\cite{Chomaz12,Corman17}. We now discuss that in the slab geometry, an asymmetry in the reponse also arises  in a model of independent scatterers  (IS), \textit{i.e.} neglecting DDI entirely. We consider an ensemble of $J=0 \to J=1$ atoms  driven by a near-resonant $\sigma^+$-polarized laser. We treat the case of normal incidence and specialize to a slab geometry of density  $\rho_{\mathrm{3D}}=\rho_{\mathrm{2D}} f(z)$, with $f(z)$ a function characterizing the precise geometry of the  slab. By symmetry, the mean electric polarization density $\overline{\bm{P}}$  and the mean (coherent) electric field depend only on $z$. The $\sigma^+$ component of the forward-scattered component of the average electric field   is then given by\,\cite{Morice95}
 \begin{align}\label{eq:Eslab}
 \overline{E}_+(z)  \ee^{\ii \kL z}  & = E_{\mathrm{L}} \ee^{\ii \kL z} +\frac{1}{\epsilon_0} \int  dz' g( z-z')   \overline{P}_{+}(z'),
\end{align} 
with $E_{\mathrm{L}} $ the amplitude of the incident probe wave and $g(z-z') = (\ii \kL/2) \ee^{\ii \kL \vert z-z'\vert}$ the Green function of the one-dimensional Helmholtz equation\,\cite{SuppMat}. A continuous medium of IS corresponds to a dielectric susceptibility $\chi = \alpha_0 \rho_{\mathrm{3D}}$, with $\alpha_0 = -6\pi \kL^{-3}/(\widetilde{\delta}+\ii)$ the single-atom electric polarisability. Substituting the polarisation $ \overline{P}_+(z) = \varepsilon_0 \chi \overline{E}_+(z)\ee^{\ii \kL z}$ in Eq.\,(\ref{eq:Eslab}), we obtain a closed integral equation
\begin{align}
\label{eq:Eparallel}
\overline{E}_+ (z)  &= E_{\mathrm{L}} - \frac{3 \ii \pi \widetilde{\rho}_{\mathrm{2D}}  }{  \widetilde{\delta}+\ii} \Big(    \int_{-\infty}^{z} d z' \,f(z')\overline{E}_+ (z')  \\
\nonumber
& \hspace{1.5cm}+ \int_{z}^{+\infty} dz' \,f(z') \overline{E}_+ (z') \ee^{-2\ii \kL(z -z')}\Big).
\end{align}
The contribution to the field at $z$ from the downstream region $ z'>z $ comes from coherently reflected (or backwards-scattered) waves with amplitude $\propto \ee^{\ii \kL z'} \ee^{\ii \kL \vert z-z' \vert} = \ee^{\ii \kL z}\ee^{-2\ii \kL (z-z')}$. The first factor is due to the incident wave and the second one to the spherical wave emitted by a dipole at $z'$. In contrast, waves transmitted through the upstream region $z'<z$  have amplitudes $\propto \ee^{\ii \kL z'} \ee^{\ii \kL \vert z-z' \vert}=\ee^{\ii \kL z}$ and thus always add up  in phase irrespective of the locations of the dipoles. 

We now specialize to the experimentally relevant ``Gaussian slab'' geometry. When $\kL \Delta z\gg 1$ (a thick sample in the ``3D regime''), the  coherent addition of reflections from the downstream region  blurs to a negligible contribution. 
Neglecting the last term in Eq.\,(\ref{eq:Eparallel}), we recover after some algebra~\cite{SuppMat} the standard formula for the complex transmission of a coherent beam through a dilute 3D sample,
\begin{align}
\label{eq:BL}
t_\mathrm{BL} = \frac{\overline{E}_+ (z\to\infty)}{E_{\mathrm{L}} }\simeq \exp \left(-\frac{ 3\pi \widetilde{\rho}_{\mathrm{2D}}(1+\ii \widetilde{\delta})  }{ 1+\widetilde{\delta}^2} \right).
\end{align}
This formula is usually derived from the Helmholtz equation by making a slowly-varying-envelope approximation, which is indeed justified for a thick sample. The intensity transmission obeys the well-known Beer-Lambert (BL) law $T_{\mathrm{BL}}=\vert t_{\mathrm{BL}} \vert^2 = \exp\left[- 6\pi \widetilde{\rho}_{\mathrm{2D}}/(1+\widetilde{\delta}^2)\right]$, and we will refer to Eq.\,(\ref{eq:BL}) as ``generalized BL law'' in the following.

\begin{figure}[b]
\includegraphics[ width=\columnwidth]{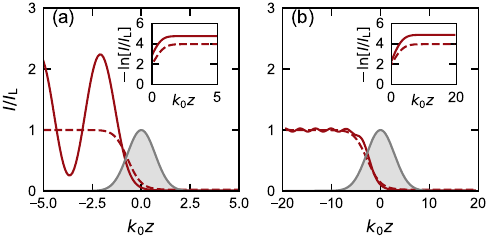}
\caption{Calculated intensity transmission in the model of independent scatterers as a function of propagation distance $z$ for (a) $\kL \Delta z = 1/\sqrt{2}$ and (b) $2 \sqrt{2}$, and  $\widetilde{\rho}_{\mathrm{2D}}=0.5$. The dashed lines show the  generalized Beer-Lambert formula $T_{\mathrm{BL}}$. The grey areas show the Gaussian density distribution $f(z)$ of the atomic slab.  }
\label{Fig:Fig4}
\end{figure} 

When $\kL \Delta z$ is comparable to or smaller than unity, the slowly-varying-envelope approximation breaks down. To describe this regime, we solve the integral equation (\ref{eq:Eparallel}) numerically. The calculated intensity profiles are shown in Fig.\,\ref{Fig:Fig4}(a-b). For a thin slab with $\kL\Delta z<1$, the intensity deviates significantly from the Beer-Lambert law. A substantial reflected component appears and interferes with the incident wave to produce the large-amplitude oscillations for $z<0$. In the limit of an infinitely thin sample  where $f(z)$ tends to a Dirac distribution,
the complex transmission  $t_\mathrm{2D} =(\widetilde{\delta}+\ii)/[\widetilde{\delta}+\ii(1+ 3\pi \widetilde{\rho}_{\mathrm{2D}})]$ no longer obeys the generalized BL law. Outside the 3D or 2D limits, the IS model with $\kL \Delta z \gtrsim 1$ exhibits asymmetric profiles in agreement with the experimental data (see Fig.\,\ref{Fig:Fig2}). We interpret the asymmetry as an etalon effect in an inhomogeneous medium, where the  reflected (backscattered) wave is small but not negligible and interferes with the incident or transmitted (forward scattered) waves.

The importance of DDI depends on the dimensionless 3D density $\rho_\mathrm{3D}\kL^{-3} \propto \widetilde{\rho}_\mathrm{2D}/(\kL \Delta z)$ and therefore increases with $\widetilde{\rho}_\mathrm{2D}$.  Following Ref.~\cite{Chomaz12,Corman17}, we perform numerical simulations of the CD model to gauge the importance of DDI in our experiments~\cite{SuppMat}).
We compare the experimental data in Figs.\,\ref{Fig:Fig2} and \ref{Fig:Fig3}  to  the three models discussed so far, the  generalized BL model in Eq.\,(\ref{eq:BL}), the IS model taking the quasi-2D geometry into account, and the CD model also including DDI. Quantitatively, the generalized BL model  overestimates both the transmission minimum and the phase shift amplitude in  Fig.\,\ref{Fig:Fig3}  by about 20\,\% (and also does not account for the asymmetric lineshapes in Fig.\,\ref{Fig:Fig2}). The IS model is slightly closer to the data, but the correction is too small to explain the experiments quantitatively. Only the CD model, shown as green dotted lines in Figs.\,\ref{Fig:Fig2} and \ref{Fig:Fig3}, can explain all the observed features, reproducing well (without any free parameter) the lineshapes in  Fig.\,\ref{Fig:Fig2} and the amplitudes in Fig.\,\ref{Fig:Fig3} within the experimental uncertainties. Note, however, that these uncertainties are comparable with the difference between the data and the less accurate (BL) model. Moreover, the saturation parameter $I/I_{\mathrm{sat}} \sim 0.2$ is not negligibly small, and the possibility that the observed reduction of the optical response stems (at least partially) from saturation effects cannot be excluded.

In previous experimental work~\cite{Jennewein16,Corman17,
Jennewein18}, the intensity transmission has been recorded directly rather than inferred from the fringe contrast. We are able to perform the same measurement in our setup since the (so far unused) ``background'' intensity $A$ fitted to the measured profiles [see Eq.\,(\ref{eq:Ifit})] actually corresponds to $ I_{\mathrm{probe}} +I_{\mathrm{L}}$, with $ I_{\mathrm{probe/L}}$ the intensities of the reference/probe waves\,\footnote{We have verified that we obtain the same results in a direct measurement where we block the reference arm of the interferometer}. 
We find that the  apparent optical depth $D_\mathrm{app} = -\ln  I_{\mathrm{probe}} / I_{\mathrm{L}} $ displays a significantly broadened lineshape and saturates near the value $D_\mathrm{sat}\approx 2.35 $\,\cite{SuppMat}. Both observations agree with earlier results obtained in the same geometry\,\cite{Corman17}, but are apparently in conflict with our fringe-contrast-based measurements and with the IS or CD models. 

To resolve this conflict, we consider the role of \textit{off-axis} scattering, \textit{i.e.} photons scattered from the incident mode $\bm k_\mathrm{L}$ towards modes $\bm k$ almost (but not quite) collinear with $ \bm k_\mathrm{L}$, and of imaging noise. Photons scattered to off-axis modes are collected by the imaging system in a cone of half-angle $\Theta \simeq R/f\approx0.35$, with $R$ and$f$ the radius and focal length of the front lens of the imaging objective. Their contribution is effectively filtered out in the interferometric scheme, but not in the direct transmission measurements where they contribute to the experimental signal. Imaging noise, dominated by the almost spatially uniform read-out noise of the camera, also contributes at high extinction levels. A simple model presented in the Supplemental Material~\cite{SuppMat} allows us to write the apparent optical depth as
\begin{equation}
\label{Eq:Texp}
D_{\mathrm{app}}  \simeq -\ln\left[  T+(1-T)\left( \frac{\Theta^2}{4}+T_{\mathrm{noise}}\right)\right].
\end{equation} 
The first term $T$ is the contribution from forward-scattering considered thus far, the second term $\propto \Theta^2$ is the contribution from off-axis scattering, $T_{\mathrm{noise}}$ is the contribution from imaging noise (dominated by the camera read-out noise) and the factor $1-T$ accounts for the ``shadow'' effect caused by the extinction of the forward-scattered component as it propagates through the slab. As the surface density increases and $T \to 0$, the measured optical depth saturates to the value $-\ln\left( \Theta^2/4+T_{\mathrm{noise}}\right) \approx 2.26$, in good agreement with the observed value \textit{without any free parameters}. A saturated lineshape also explains the observed line broadening following the argument given in \cite{Guerin17}.

In summary, we have measured the complex optical response of a quasi-2D slab of ultracold atoms with random positions. Our experimental results settle the questions raised in previous works about the relative roles of dipole-dipole interactions, geometry and two- versus multi-level character of the atomic transition~\cite{Jennewein16,Bromley16,Corman17,Jennewein18}. In particular, we find that the discrepancies between experiments and theory noted in earlier works can be resolved by accounting for off-axis scattering and imaging noise. Our experimental data are compatible with the predictions of  numerical simulations for coupled, frozen dipoles. 
By increasing the confinement to the 2D regime $\kL \Delta z \ll 1$, one should access a regime of optically dense 2D systems with strong DDI considered to realize quantum memories~\cite{Jenkins12,Ferioli21}, single-layer atomic mirrors~\cite{Rui20,Srakaew23}, or to investigate topological optics~\cite{Bettles16,Perczel17,Shahmoon17}. Whether or not the CD model remains sufficient to describe the atomic response  is an open question, as DDI may eventually trigger spatial dynamics inside  the cloud when $\rho_{\mathrm{3D}} \kL^{-3} \geq 1$.  

\begin{acknowledgments}
We thank Tristan Villain for his contribution in the early stage of this project and Jean Dalibard for stimulating discussions. We are also grateful to Yvan Castin,  Pierre Clad\'e and F\'elix Werner for their assistance with the numerical simulations.
\end{acknowledgments}

%

\bibliography{ResonantLightScattering}
\bibliographystyle{apsrev4-2}
\definecolor{violet3}{RGB}{120,0,180}
\hypersetup{urlcolor=violet3}


\end{document}